# Voids in Real Space and in Redshift Space

Barbara S. Ryden [1]

Department of Astronomy, The Ohio State University, Columbus, OH 43210;

ryden@mps.ohio-state.edu

Adrian L. Melott

Department of Physics & Astronomy, University of Kansas, Lawrence, KS 66045;

melott@kusmos.phsx.ukans.edu

## ABSTRACT

Using two-dimensional numerical simulations of gravitational clustering, with initial power spectra of the form $P(k) \propto k^n$, we compare the properties of voids in real space to their properties in redshift space. Both the void probability function (VPF) and the underdense probability function (UPF) are enhanced in redshift space. The enhancement is greatest in the limit $n = -2$. The VPF and UPF treat voids statistically and assume that two-dimensional voids are circular; we present an algorithm which detects individual voids. Voids found by this algorithm are ellipses whose enclosed density of galaxies falls below a threshold density. When voids are identified using this algorithm, the mean void size and the maximum void size both increase in going from real space to redshift space. The increase is greatest in the limit $n = -2$. In redshift space, the principal axes of the largest voids in the $n = 2$ and $n = 0$ simulations show a statistically significant tendency (at the 95% confidence level) to be distributed anisotropically, relative to the line of sight from the origin to the void center.

*Subject headings:* cosmology: theory – galaxies: distances and redshifts – large-scale structure of universe

## 1. Introduction

When the positions of galaxies are mapped in redshift space, the large-scale structure which they trace is seen to be "frothy" or "bubbly" in nature. Much of the map is filled with voids – regions where there are few or no galaxies. The galaxies are restricted mainly to thin filaments or sheets which lie between the voids. Comparing the observed pattern of voids with the pattern predicted by a given model of structure formation is potentially a very powerful test for accepting or rejecting that model. Unfortunately, the differences between models are often subtle. A wide range of possible models for the formation of structure predict a bubbly distribution of galaxies. In the gravitational instability scenario, regions which are originally underdense initially expand more rapidly than the Hubble flow (Fillmore & Goldreich 1984; Bertschinger 1985), with the excess expansion rate dependent upon the initial density profile. As the underdense region expands, it becomes more nearly spherical (Fujimoto 1983; Icke 1984; Blaes, Goldreich, & Villumsen 1990; Ryden 1994). As the present large scale structure reveals, voids are no longer isolated spherical regions, but form a space-filling foam. Gott, Melott, and Dickinson (1986) – see also Melott (1990) – emphasized the "spongelike" topology of a system of superclusters and voids evolved from Gaussian initial conditions. Both voids and superclusters interlock. Certain kinds of shifts from initial conditions due to nonlinear evolution are detectable. More recently, Sahni et al. (1994) have looked at substructure in voids and noted that it is generic, often filamentary, and evolves by progressively emptying. Yess and Shandarin (1995) have used enhanced percolation techniques to find that there is a "bubble shift" toward isolated voids for initial conditions with more power on large scales.

---

[1] National Science Foundation Young Investigator

If structure grows via gravitational instability, then the size and shape of voids in real space depends on the initial power spectrum $P(k)$ of density fluctuations and on the density parameter $\Omega$. In an open universe, void evolution stops when $\Omega \sim 0.5$; thereafter, the void network simply expands along with the Hubble flow (Regös & Geller 1991). In an $\Omega = 1$ universe with a power-law spectrum, $P(k) \propto k^n$, the void network evolves with time in a self-similar manner (Beacom et al. 1991; hereafter BDMPS). At any epoch, the mean void radius is proportional to the nonlinearity scale $2\pi/k_{n\ell}$ (Kauffman & Melott 1992). The spectrum of void sizes, however, depends on the initial power-law index $n$. The range of void sizes can also distinguish between a universe dominated by hot dark matter (HDM) and one dominated by cold dark matter (CDM). In simulations normalized to match the observed correlation function of galaxies, voids in an HDM simulation are twice the size of voids in a CDM simulation (Melott 1987).

If we knew with absolute accuracy the position of galaxies in real space, we could use the size of voids to place constraints upon the initial power spectrum $P(k)$. However, measuring the physical distances to galaxies is difficult; it is much easier to measure their redshifts. Consequently, practical studies of the properties of voids must take place in redshift space rather than real space. If all galaxies were smoothly following the Hubble flow, with no peculiar velocities, and if the Hubble constant $H$ were truly constant with time, then the mapping between real space and redshift space would be linear. In the general case, however, the Hubble constant changes with time, and structures which are isotropic in real space become distorted in redshift space when the redshift $z$ is large (Alcock & Paczyński 1979) In fact, distortions in redshift space can provide an estimate of the value of the deceleration parameter $q_0$ (Phillips 1994; Ryden 1995). However, the distortions which result from the changing value of $H$ only become large when $z \gtrsim 1$. In the nearby universe, where $z < 1$, the dominant contribution to distortions in redshift space comes from the peculiar velocities of galaxies.

The goal of this paper is to compare the properties of voids in redshift space with the properties of voids in real space. We consider galaxies at $z \lesssim 0.2$; since we are dealing with nearby galaxies, we will only include the effects of peculiar velocity distortions. The effects of cosmological distortions, occurring when $q_0 > -1$, will be dealt with in a future paper (Ryden & Melott 1995).

To study systematically the effect of the initial power spectrum, we will use simulations with power-law initial spectra, $P(k) \propto k^n$. To obtain a large dynamic range, we use the two-dimensional simulations of BDMPS; these simulations can be analyzed in the same manner as a "slice" redshift survey. The extent to which voids are distorted by peculiar velocities depends not only on the underlying distribution of galaxies but also on how we define voids in the galaxy distribution. Some of the statistics used to measure voids define a "void" as a region which is totally empty of galaxies. The void probability function (or VPF) is one such statistic. The VPF $P_0(a)$ is the probability that a randomly positioned circle of area $a$ contains no galaxies (White 1979). (In this paper, we will deal with two-dimensional surveys; in a three-dimensional galaxy survey, the VPF $P_0(V)$ is the probability that a sphere of volume $V$ contains no galaxies.)

However, from a physical standpoint, it makes more sense to define a void as a region which is underdense when compared to the average number density of galaxies. Little & Weinberg (1994) use the underdense probability function (or UPF) $P_{80}$, which they define as the probability that a randomly positioned circle of area $a$ (or sphere of volume $V$) has an average density more than 80% below the global mean density. The UPF has the advantage of being relatively insensitive to the sparseness with which the galaxies are sampled. The underdensity level of $\delta \rho / \rho = -0.80$ is chosen to match the density contrast of the largest voids in the CfA redshift survey (Vogeley, Geller, & Huchra 1991; Vogeley et al. 1994).

In addition to statistical measurements such as the VPF and UPF, there exist algorithms for identifying individual voids within a sample (Kauffmann & Fairall 1991; Kauffman & Melott 1992; Ryden 1995). These void-detection algorithms, in addition to providing a spectrum of void sizes, also enable us to specify the shapes and positions of voids. In this paper, we use an algorithm based on that of Ryden (1995) which identifies voids as non-overlapping ellipses which are either totally empty of galaxies or with an average density within the ellipse which is 80% below the mean density. Since the voids, by this definition, are not constrained to be circular, we can study the axis ratios and orientations of the elliptical voids. This enables us to find out whether the net effect of peculiar velocities is to elongate voids along the line of sight or to squash them along the line of sight.

## 2. The Simulations

We calculate the properties of voids in a set of two-dimensional computer simulations of gravitational clustering. These simulations are described in detail in BDMPS; we will only give a brief description here. The simulations are designed to solve the Poisson equation for two-dimensional density perturbations in a three-dimensional $\Omega = 1$ matter-dominated universe. The use of 2D simulations rather than 3D allows higher spatial resolution for a given number of mass particles. The study of 2D simulations can give useful insights into the behavior of a fully three-dimensional universe. The two-point correlation function (BDMPS), the three-point correlation function (Fry, Melott, & Shandarin 1992), the phase correlations in Fourier space (Scherrer, Melott, & Shandarin 1991), the spectrum of void sizes (Kauffmann & Melott 1992), and the percolation of overdense regions (Dominik & Shandarin 1992) have all been studied for this set of simulations. Their evolution can be observed in the videotape accompanying the *ApJ* for Kauffmann and Melott (1992).

The simulations were done with a particle-mesh code using $512^2$ particles on a $512^2$ mesh. The initial power spectra were of the form $P(k) \equiv <|\delta_k|^2> \propto k^n$, with $n = 2$, $n = 0$, or $n = -2$. The low-frequency cutoff was at the fundamental frequency $k_f = 2\pi/L$, where $L$ is the length of one side of the simulation box. The high-frequency cutoff was at the Nyquist frequency $k_{Ny} = 256 k_f$. We analyzed each model at the timestep when the scale of nonlinearity is $k_{n\ell} = 32 k_f$. The value of $k_{n\ell}$ is computed from the relation

$$(\delta\rho/\rho)^2 = \int_0^{k_{n\ell}} |\delta_k|^2 \, d^2k = 1 \; , \tag{1}$$

where the Fourier transforms $\delta_k$ of the overdensity are computed by linear extrapolation from the initial conditions.

The density fluctuations in the real universe do not have a pure power-law spectrum. A two-dimensional spectrum with $n = 2$ corresponds to $n = 1$ in three dimensions; this is the Harrison-Peebles-Zel'dovich spectrum predicted on large scales by inflationary scenarios. A two-dimensional $n = 0$ spectrum is analogous to $n = -1$ in three dimensions; this is approximately the slope of the observed power spectrum on the scale of galaxy clusters. A two-dimensional $n = -2$ spectrum corresponds to $n = -3$ in three dimensions; this is the slope predicted on small scales for a CDM dominated universe. The $n = -2$ case is somewhat pathological, since the mean square overdensity diverges at small wavenumbers. Thus, an $n = -2$ simulation will have structure on all scales up to the box size $L$, and will be strongly affected by the periodic boundary conditions of the simulations. Nevertheless, we will examine the $n = -2$ simulations as an example of a universe with power weighted toward large scales.

To compare simulations to reality, we first reduced the number of particles by randomly selecting one out of every four mass points and designating them "galaxies". In other words, we assumed unbiased galaxy formation. The effects of biasing on the VPF and UPF are discussed by Little & Weinberg (1994), who find that the shape and amplitude of the VPF depend on the biasing scheme adopted. In view of the uncertainty in how biasing (if any) works, we look only at the unbiased distribution. In making redshift plots, we adopt a box size of $L = 0.38$ in redshift units, corresponding to $L = 1140h^{-1}$ Mpc in physical units, where the Hubble constant is $H_0 = 100h$ km sec$^{-1}$ Mpc$^{-1}$. This scale is chosen so that the rms overdensity of galaxies within a circle of radius $r_1 = 8h^{-1}$ Mpc is equal to one for the $n = 2$ and $n = 0$ simulations. With this normalization, we have built in a characteristic area for the large scale structure, equal to

$$a_S = \pi(H_0 r_1/c)^2 = 2.2 \times 10^{-5} \tag{2}$$

in redshift units, or $a_S = 200h^{-2}$ Mpc$^2$ in physical units. With the normalization chosen in this way, the mean surface density of galaxies is equal to $\Sigma = 256^2/(0.38)^2 = 4.5 \times 10^5$ in redshift units, or $\Sigma = 0.050h^2$ Mpc$^{-2}$ in physical units. The characteristic discreteness scale is thus

$$a_D \equiv \Sigma^{-1} = 2.2 \times 10^{-6} = 20h^{-2}\,\mathrm{Mpc}^2 \ . \tag{3}$$

The discreteness scale $a_D$ is comfortably smaller than the scale $a_S$ despite the reduction by a factor of four in the number of particles (a reduction which was performed for numerical convenience in computing the void statistics).

The galaxy distributions for the $n = 2$, $n = 0$, and $n = -2$ simulations are shown in Figure 1(a), 2(a), and 3(a), respectively. Distances on the plot are given in redshift units; to convert to physical units, multiply by $c/H_0 = 3000h^{-1}$ Mpc. In recognition of the fact that the real universe doesn't have periodic boundary conditions (at least not on the scale of a gigaparsec), the simulations are plotted in a form reminiscent of a volume-limited redshift survey, with only those particles at a distance less than $z = 0.19$ from an arbitrarily chosen origin being shown. If a Flatland astronomer located at the origin of Figures 1(a), 2(a), or 3(a) performs a redshift survey, though, the measured redshift for will contain a component from the radial component of the peculiar velocity of each galaxy. The resulting plots in redshift space of each of the three simulations are given in Figures 1(b), 2(b), and 3(b), for $n = 2$, $n = 0$, and $n = -2$ respectively. Only galaxies with a redshift $z < 0.19$ are shown.

## 3. Statistics of Voids

The void probability function (White 1979; Fry 1985, 1986; Otto et al. 1986), has been the most widely used statistic for measuring the size of voids, having been applied to redshift surveys of galaxies (Vogeley et al. 1991, 1994; Fry et al. 1989; Einasto et al. 1991) and to numerical simulations (Fry et al. 1989; Einasto et al. 1991; Weinberg & Cole 1992; Little & Weinberg 1994; Vogeley et al. 1994; Ghigna et al. 1994).

Other statistics that have been used can be related to the VPF. For instance, Ryden & Turner (1984) measured the distance $r$ from positions within a survey to the nearest neighboring galaxy; the distribution $P_{nn}(r)$ of the nearest neighbor distances is related to the VPF $P_0(\pi r^2)$ by the relation

$$P_{nn}(r) = -\frac{dP_0}{da}\frac{da}{dr} = -2\pi r \frac{dP_0}{da} \ . \tag{4}$$

The void probability functions for the computer simulations are shown in the left hand panels of Figure 4. The solid line gives the VPF in real space, and the dotted line gives the VPF in redshift space. In each panel, the dashed line gives the VPF for a Poisson distribution with the same surface density $\Sigma = 4.5 \times 10^5$ as the computer simulations. For a Poisson distribution of points, the VPF is

$$P_0(a) = e^{-a/a_D} , \qquad (5)$$

where $a_D = \Sigma^{-1}$ is the mean area per particle. At the discreteness scale $a_D = 2.2 \times 10^{-6}$, the VPF for the Poisson distribution has fallen to $P_0(a_D) = 1/e \approx 0.37$. Because of small scale clustering, the VPFs for the numerical simulations are larger at this scale, ranging from $P_0(a_D) = 0.52$ for the $n = -2$ simulation (in real space) to $P_0(a_D) = 0.76$ for the $n = 2$ simulation. In general, for a fixed value of $a$ the amplitude of $P_0(a)$ increases as the spectral index $n$ increases.

Moving from real space to redshift space changes the shape of the VPF. There is only a small effect when $n = 2$ (upper left panel in Figure 4); in this simulation, the concentration of power on small scales means that the distortions in redshift space take the form of short "fingers of God", which do not strongly affect the size of the empty regions. When $n = 2$, the VPF in redshift space is slightly decreased when $a \lesssim a_S$ and slightly enhanced when $a \gtrsim a_S$. On small scales, the fingers of God protrude into the voids in redshift space, thus decreasing the VPF (Vogeley et al. 1991); on large scales, bulk flows increase the amplitude of density fluctuations in redshift space, thus increasing the VPF (Kaiser 1987).

The most obvious change in the VPF occurs when $n = -2$. In this case, the presence of power on large scales results in bulk flows which enhance the probability of large empty voids. In real space, the largest empty void in the $n = -2$ simulation has $a \approx 33 a_D \approx 3.3 a_S$. In redshift space, the largest empty void has $a \approx 120 a_D \approx 12 a_S$. Three-dimensional simulations also show a spectrum-dependent increase of the VPF on large scales in redshift space; the enhancement of the VPF also depends on the density parameter $\Omega$, the cosmological constant $\Lambda$, the bias parameter $b$, and the bias mechanism chosen (Little & Weinberg 1994).

Historically, the VPF has been the most commonly used statistic for measuring the voids in redshift surveys and in simulations. It has long been recognized, however, that the VPF has serious shortcomings. It is strongly dependent, for instance, on the number density $\Sigma$ of galaxies. If the VPF is naïvely applied to a magnitude-limited survey, in which the number density of observed galaxies falls off with redshift, it will tell us (erroneously) that the characteristic void size increases with redshift. Furthermore, the discovery of one new galaxy in the center of a void can drastically change the result – an unfortunate sensitivity to completeness of surveys. What we really want is a statistic which is not sensitive to the sampling efficiency of our survey. A more sensible alternative to the VPF is the underdense probability function, or UPF, in which a void is defined as a region whose number density of galaxies falls below a threshold density $f\Sigma(z)$, where $0 < f < 1$. (Note that we are now permitting the surface density $\Sigma$ to be a function of redshift.) Using this definition, we see that the VPF is simply the limiting case of the UPF with $f = 0$. Since substructure in voids is generic in gravitational clustering (Sahni et al. 1994), we prefer $f \neq 0$.

– 7 –

We now have the freedom to set the threshold $f$ in the UPF. Following previous work (Vogeley et al. 1991; Weinberg & Cole 1992), we set $f = 0.2$ and compute the UPF $P_{80}(a)$ – we are using the notation of Weinberg & Cole, in which the UPF computed with a threshold $f$ is designated as $P_m(a)$, with $m \equiv 100(1-f)$. For each of the circular surveys in Figures 1, 2, and 3, we began by computing the number density $\Sigma(z)$. For illustrative purposes, we simply found the best-fitting polynomial of the form $\Sigma(z) = c_0 + c_1 z + c_2 z^2$, since $\Sigma$ in these simulations is nearly constant with $z$. In a real redshift survey, a more appropriate choice might be to do a nonparametric fit to the number density (Merritt & Tremblay 1994). We then found the probability that a circle of area $a$ centered at a redshift $z$ contains a number $N$ of galaxies such that $N/a < f\Sigma(z)$. This probability, expressed as a function of $a$, is the UPF $P_m$. For areas $a < 1/[f\Sigma(z)]$, the UPF and the VPF are identical, since a void with such a small area can only fall below the density threshold if it contains $N = 0$ galaxies. Circles with an area $a > M/[f\Sigma(z)]$ will be underdense if they contain fewer than $M$ galaxies. If $\Sigma$ is constant, the UPF will show discreteness effects from shot noise. In a Poisson distribution, for example,

$$P_{80}(a) = e^{-a/a_D} \sum_{n=0}^{n_{max}} \frac{(a/a_D)^n}{n!} , \qquad (6)$$

where $n_{max} = \text{int}(0.2a/a_D)$, and $a_D = \Sigma^{-1}$. The discreteness effects are insignificant only when $a \gg f^{-1} a_D$.

The UPF for our numerical simulations is shown in the right hand panels of Figure 4. The solid line in each panel is the UPF calculated in real space; the dotted line is the UPF calculated in redshift space. The mean number density $\Sigma(z)$ for each simulation is nearly constant as a function of $z$; hence the discreteness effects are easily seen, with the UPF jumping upward whenever $a$ is an integral multiple of $f^{-1} <a_D> = 1.1 \times 10^{-5}$, where $<a_D> = 2.2 \times 10^{-6}$ is the discreteness scale averaged over the entire mock survey, and the chosen threshold is $f = 0.2$. The dashed line in each panel is the UPF calculated for a Poisson distribution with $a_D = 2.2 \times 10^{-6}$ and $f = 0.2$.

In real space, the amplitude of the UPF at small areas $(a < f^{-1} <a_D>)$ increases as the index $n$ increases; in this range of areas, as mentioned earlier, the UPF and the VPF are identical. At larger areas $(a \gtrsim 8f^{-1} <a_D>)$, the amplitude of the UPF *decreases* as $n$ increases. In particular, the $n = -2$ simulation has a tail of underdense voids reaching to a maximum void size of $a \sim 3 \times 10^{-3} \sim 1400 <a_D>$. The effect of peculiar velocities is to enhance the amplitude of the UPF at large void areas.

The moral of the story, in summary, is that the frequency of voids in a simulations depends on how a void is defined. The VPF, in which a void is defined as a totally empty region, is lower in amplitude than the UPF, in which a void is defined as an underdense region. The difference between the VPF and UPF is particularly great in the limit $n \to -2$ ($n \to -3$ in three dimensions). Moreover, the frequency of voids, as measured by either the VPF or UPF, is different in redshift space than in real space. For voids larger than the characteristic discreteness scale, the amplitude of the VPF or UPF in redshift space is higher than the amplitude in real space. The difference between redshift space and real space is particularly great, once again, when $n \to -2$.

## 4. Defining Individual Voids

The UPF and the VPF give a statistical measure of how many voids are in a given distribution, and how large those voids are. For many purposes, however, it is useful not merely to compute statistics, but to identify individual voids. For each void, we then know the void's location, area, shape, and orientation. Kauffmann & Fairall (1991) used an algorithm which defined a void as a region empty of galaxies to compile a catalog of voids in the galaxy distribution. Kauffmann & Melott (1992) used a similar algorithm, but now defining a void as an underdense region, to locate the voids in numerical simulations. The algorithms used by Kauffmann and her collaborators start by finding the largest square which fits within a void. This square, called the base square, gives a first approximation to the size of the void. Supplementary rectangles are then added to each edge of the square in order to approximate more closely the true size and shape of the void. This algorithm gives a good fit the the shapes of circular voids. However, if the true void is an ellipse whose principal axes lie along the diagonals of the base square, its shape is relatively poorly approximated. Ryden (1995), attempting to measure the shapes of voids more accurately, devised an algorithm which defines a void as an ellipse rather than as a sum of rectangles. In this paper, we are interested in the shape and orientation of voids, in addition to their areas, so we will adopt, adapt, and improve the void-finding algorithm of Ryden (1995).

An additional concern is that our algorithm, as well as that of Kauffmann and Melott (1992) treats voids as isolated objects. Yet, voids evolved from Gaussian initial conditions will be interconnected to some degree (depending on the initial conditions and density threshold). Kauffmann and Melott (1992) confirmed that in extremely sponge–like topologies their method finds a size equal to "tunnel diameters". Also, Yess and Shandarin (1996) find that for three–dimensional spectra $n \leq -1$ nonlinear dynamics shift voids toward isolation. (Their work, based on percolation techniques, is able to study voids and superclusters separately. They find that gravity shifts superclusters toward connectivity for all spectra they studied).

When applying the VPF or UPF to a two-dimensional simulation, it is customary to assume that voids are circular. This is not always a valid assumption, particularly when working in redshift space. It is true that an isolated void becomes more spherical with time, but this is only true in real space; in redshift space, the void's expansion velocity causes it to be stretched out along the line of sight; this effect is well illustrated in the simulated redshift maps of Regös & Geller (1991). In addition, at redshift $z \gtrsim 1$, cosmological distortions will cause further stretching along the line of sight (Alcock & Paczyński 1979).

Naïvely, then, we might expect all voids in redshift slices to be ellipses with their long axes along the line of sight away from the observer. However, there is a complicating factor which arises from the fact that voids are not isolated structures, but rather form a space-filling foam.

The tendency for a void to expand will be inhibited by the fact that it is surrounded by voids similar in size to itself. If two adjacent voids are of comparable size, then the peculiar velocities of the galaxies in the wall between the voids will be small (Melott 1983; Regös & Geller 1991; Dubinski et al. 1993). The shapes of voids in redshift space will also be altered by the presence of virialized clusters in the walls between voids. In redshift space, these clusters will form fingers of God poking into the voids and decreasing their size. Furthermore, any bulk flows which exist on scales larger than the characteristic void size will also cause distortions in redshift space.

The isolated void model suggests large peculiar velocities at the boundary of voids, which was a problem raised when early data came out. Of course, velocities will be zero in a universe filled with voids of equal size. But voids are typically not of equal size. As noted by Melott (1983), evolution of the void network proceeds by larger voids squashing smaller ones, as peculiar velocities in walls are normally directed from larger voids to smaller ones. One would then expect that the dispersion of void sizes in redshift space will be larger than in redshift space. Results on this will be shown later.

In the nearby universe ($z \lesssim 0.04$) the net effect of these peculiar velocities, at least to lowest order, is to cancel each other out. A study of the CfA redshift slice (Slezak, de Lapparent, & Bijaoui 1993) identifies 15 elliptical voids, which show no tendency to be elongated along or perpendicular to the line of sight. How severe a constraint, we may ask, does this lack of distortions place on the initial conditions in the gravitational clustering scenario? If the universe is open, with $\Omega \ll 1$, we expect negligible distortion of the void network. In a universe with $\Omega = 1$, though, it is also possible that there will be no net distortion of void shapes. With the ensemble of $\Omega = 1$ simulations that we are examining, we can find out for which values of $n$ (if any) there are detectable distortions of the large voids in redshift space.

To implement our method for identifying voids, we start by laying down a Cartesian grid of points over our simulations. For the $n = 0$ and $n = 2$ simulations, we use a grid spacing $\delta z = 1.27 \times 10^{-3}$; for the $n = -2$, which has contains larger voids, we use the larger spacing $\delta z = 1.52 \times 10^{-3}$. These grid spacings give a resolution in area of $\sim (\delta z)^2 \sim 1.6 \times 10^{-6}$ for $n = 2$ and $n = 0$, and $\sim (\delta z)^2 \sim 2.3 \times 10^{-6}$ for $n = -2$, comparable to the discreteness scale $a_D = 2.2 \times 10^{-6}$. For each grid point, we find the largest ellipse centered on that point within which the mean density of galaxies falls below a threshold density $f\Sigma(z)$, where $z$ is the distance of the test point from the origin. We investigate both the $f = 0$ case, where voids are defined as being totally empty, and the $f = 0.2$ case, where voids are underdense. The ellipses centered on a particular test point are characterized by their area $a$, their axis ratio $q$, equal to the semimajor axis divided by the semiminor axis, and their position angle $\phi$ relative to the $x$ axis of the Cartesian grid. We search through $(a, q, \phi)$ parameter space to find the ellipse with the largest value of $a$, subject to the constraint that its density fall below the threshold density. The ellipse must also lie entirely within the limits of our simulated survey, which ends abruptly at $z = 0.19$. We constrain $q$ to lie in the range $1 \leq q \leq 3$. Our upper limit on $q$ is somewhat arbitrary; any underdense region with an axis ratio greater than 3 : 1, we decree, is no longer a "void", but a "tunnel". The position angle $\phi$ is allowed to vary from 0 degrees to 180 degrees. Note that this is a refinement of the original technique of Ryden (1995), who constrained the principal axes of the ellipse to lie along and perpendicular to the line of sight to the origin. In our search through parameter space, we find the maximum area void using a step size $\Delta \ln q = 0.061$ in the axis ratio, and $\Delta \phi = 5°$ in the position angle.

After our search is complete, we have, for each grid point $(x, y)$, the area $a$, the axis ratio $q$, and the position angle $\phi$ of the largest ellipse centered on the grid point whose density falls below the designated threshold $f\Sigma(z)$, where $z^2 = x^2 + y^2$. Our task of identifying voids, however, is not yet complete, since we want our voids to be nonoverlapping regions. Our next step, then, is to take the ellipses associated with the grid points, and rank them according to decreasing area. The largest of all the ellipses is designated the largest void in the survey. The next largest ellipse is checked to see whether it overlaps with the largest void. If it does not, it is designated the second largest void. If it does overlap, though, it is discarded, and we go to the next ellipse on our list. This process is continued until we reach a void size equal to the resolution limit $(\delta z)^2$.

The voids defined by this technique are nonoverlapping ellipses, within which the density of galaxies falls below a predesignated threshold. In three dimensions, the voids found would be triaxial ellipsoids; to maximize the volume of each ellipsoid, we would have to vary the two axis ratios $(p, q)$ of the ellipsoid and the orientation $(\theta, \phi)$ of its major axis. This makes the search for voids in three dimensions much more computationally intensive.

The largest voids found in the $n = 2$ simulation, assuming a void threshold $f = 0.2$, are displayed in Figure 5. The voids in real space are shown in Figure 5(a); this plot contains 422 voids, which together fill half the total survey area. The largest void in the survey has an area $a = 3.1 \times 10^{-4}$. The voids in redshift space are shown in Figure 5(b), which displays 446 voids, which suffice to fill half the survey area. The largest void in redshift space has an area $a = 4.2 \times 10^{-4}$. Note that the voids found by our algorithm are quite elongated. Of the voids in real space shown in Figure 5(a), 80% have an axis ratio $q > 2$ (including the 12 largest voids), and 39% have an axis ratio equal to the maximum permissible value, $q = q_{max} = 3$. Of the voids in redshift space shown in Figure 5(b), 75% have $q > 2$ (including the 25 largest voids), and 31% have $q = q_{max} = 3$.

Figure 6 shows the voids found in the $n = 0$ simulation, assuming a void threshold $f = 0.2$. The results in real space are shown in Figure 6(a), which contains 585 voids, filling half the survey area. The largest void has an area $a = 4.0 \times 10^{-4}$. The results in redshift space are shown in Figure 6(b), which contains 436 voids, which together fill half the survey area. The largest void in redshift space has $a = 7.1 \times 10^{-4}$; as in the case when $n = 2$, the effect of peculiar velocity distortions is to increase the size of the largest void found in the survey.

Figure 7 shows the voids found in the $n = -2$ simulation, assuming $f = 0.2$. The voids in real space are shown in Figure 7(a); in this plot, 1335 voids are required to fill half the total survey area. The largest void has an area of $a = 3.2 \times 10^{-3}$. The effects of peculiar velocities are strongly visible in Figure 7(b), which shows the voids in redshift space for the $n = -2$ simulation. In redshift space, 184 voids are required to fill half the area, and the largest void has an area $a = 6.7 \times 10^{-3}$, more than double the size of the largest void in real space.

We can find, in different ways, a characteristic void size for the voids as we have defined them. Kauffmann & Melott (1992) found that both the mean void area and the median void area are acceptable measurements of the characteristic void size. The mean void area is computed by weighting each void's area by the fraction of the total survey area which it occupies. Thus, if there are $N_v$ voids located, the mean void area is

$$a_{\mathrm{mean}} = \sum_{i=1}^{N_v} a_i^2 \bigg/ \sum_{i=1}^{N_v} a_i \;, \tag{7}$$

where $a_i$ is the area of the $i$th largest void. The median void size, $a_{\mathrm{med}}$, is the void area such that 50% of the total survey area is contained in voids of area $a_{\mathrm{med}}$ or greater. We computed both the mean and median void sizes for each simulation.

Figure 8 shows the fraction $F(> a)$ of the total area of the survey contained in voids of area $a$ or larger. For the panels in the right hand column, voids are defined as underdense regions with $f = 0.2$. For the purpose of comparison, the panels in the left hand column show the results when voids are defined as totally empty regions ($f = 0$). Since, as discussed earlier, we believe that underdense voids are more physically meaningful than empty voids, we will only discuss below the case $f = 0.2$, and not $f = 0$. The solid line in each panel represents the distribution in real space; the dotted line represents the distribution in redshift space. On each curve, the square symbol represents the point where $a = a_{\mathrm{mean}}$; the triangular symbol represents the point where $a = a_{\mathrm{med}}$. The mean void size is larger than the median void size; the difference is particularly great when $n = -2$, and there is a wide range of void sizes present.

When $n = 2$, the characteristic void size is not strongly affected by the presence of peculiar velocity distortions in redshift space. In fact, $a_{\mathrm{mean}}$ is increased by 6% in going from real space to redshift space, while $a_{\mathrm{med}}$ is *decreased* by 24% in going from real space to redshift space. Thus, when $n = 2$ and power is concentrated on small scales, whether the characteristic void size increases or decreases in going to redshift space depends on how the characteristic size is defined. When $n = 0$, $a_{\mathrm{mean}}$ is increased by 58% in going from real space to redshift space, while $a_{\mathrm{med}}$ is also increased, but only by 7%. In the pathological case $n = -2$, both the mean and median void size are greatly increased by going from real space to redshift space; $a_{\mathrm{mean}}$ is increased by 270% and $a_{\mathrm{med}}$ is increased by 470%.

It is also interesting to examine the dispersion in $a$; these quantities are shown for the three spectra in Table 1. The data are consistent with the idea that voids are larger in redshift space. If the model of large voids squeezing smaller ones is correct, the dispersion should *increase* in redshift space. The data in the table support that hypothesis. It has been noted informally from previous simulations that initial conditions with relatively more power on large scales (small $k$) show a greater dispersion in void size. Furthermore, Kauffman and Melott (1992) did numerical experiments similar to those seen here, except that all initial power was removed for $k < k_{cut}$. The resulting distribution of voids is highly uniform in size.

We propose that the morphological expression of $\delta\rho/\rho$ on quasilinear scales is dispersion in void size. Since void diameters are close to $\lambda_{n\ell} \equiv 2\pi/k_{n\ell}$ (Kauffmann and Melott 1992) then power on scale $k < k_{n\ell}/2$ should contribute almost entirely to void size dispersion. If $k_{n\ell}/2$ is substituted for the upper bound in equation (1), the resulting $\delta\rho/\rho$ compares favorably with the last column in Table 1. A more detailed comparison would require postulating a specific window function, but the trend is clearly correct. We propose that void size dispersion may be an additional estimator of integrated power. As luminous and dark matter are unlikely to be separated on scales larger than the void size, it may be unbiased. Our conclusions on the effect of large scale power on the size and dispersion of void size disagree with those of Frisch et al. (1995). This is most likely due to their different definition of voids, corresponding to completely empty regions.

Going from real space to redshift space can thus affect the size of voids. The size of the largest void in each survey is greater in redshift space, as is the weighted mean void size. But does going from real space to redshift space significantly affect the shape of voids? A void which is expanding in comoving coordinates will be elongated along the line of sight in redshift space; a void which is shrinking in comoving coordinates will be compressed along the line of sight in redshift space. A simple visual comparison of Figure 5(a) with 5(b), 6(a) with 6(b), and 7(a) with 7(b) does not reveal a strong tendency for voids to be stretched along the line of sight, or perpendicular to the line of sight, in redshift space.

A quantitative analysis of the distortion of voids can be made by measuring the angle $\Phi$ between the long axis of each elliptical void and the line of sight between the observer at the origin and the void center. If the large scale structure is truly isotropic, then the distribution of the angle $\Phi$ for the individual voids in a sample should be statistically indistinguishable from a uniform distribution over the range $0° < \Phi < 90°$. If voids are systematically elongated along the line of sight, then the distribution will be weighted toward $\Phi < 45°$. If voids are compressed along the line of sight, then the distribution will be weighted toward $\Phi > 45°$. To test the hypothesis that large voids are isotropic, we computed the value of $\Phi$ for all underdense elliptical voids in each survey with $a > a_{\text{med}}$. We used a Kolmogorov-Smirnov (KS) test to compare the computed cumulative distribution function of $\Phi$ with the theoretically expected value

$$P(< \Phi) = (\Phi/90°) \tag{8}$$

for an isotropic distribution of voids.

As measured by a KS test, the voids in real space are consistent with the hypothesis that the position angles $\Phi$ are isotropically oriented. Specifically it estimates the probability that the observed sample could be drawn from a population of isotropically oriented voids. When $n = 2$, the KS probability is $P_{\text{KS}} = 0.146$; when $n = 0$, $P_{\text{KS}} = 0.704$; when $n = -2$, $P_{\text{KS}} = 0.429$. Since we expect the distribution of voids to be isotropic in real space, these large values for $P_{\text{KS}}$ are expected. In redshift space, the KS probabilities are smaller, but are still insufficient to rule out the hypothesis of isotropy at a high confidence level. In redshift space, when $n = 2$, the KS probability is $P_{\rm KS} = 0.044$; when $n = 0$, $P_{\rm KS} = 0.047$; when $n = -2$, $P_{\rm KS} = 0.087$. At a 95% confidence level, the orientations of large voids in the $n = 2$ and $n = 0$ differ from an isotropic distribution. The mean position angle $\Phi$ for voids with $a > a_{\rm med}$ in redshift space is 44.4° when $n = 2$ and 47.0° when $n = 0$. The mean position angle for the $n = -2$ simulation, which has a smaller statistical significance for its anisotropy, is 42.2°. Thus, in the $n = 0$ simulation, voids tend to be compressed along the line of sight, while in the $n = -2$ simulation, they tend to be elongated along the line of sight.

## 5. Conclusions and Speculations

Since galaxy redshift surveys create maps in redshift space and not in real space, it is useful to understand how the properties of voids in redshift space differ from their properties in real space. Historically, the VPF and later the UPF have been the most prominent tools for measuring the properties of voids. For a two-dimensional simulation with $n = 2$, the power is concentrated on small scales, and the most prominent distortions in redshift space are short fingers of God. These have a relatively small effect on the VPF and UPF in redshift space, causing a small decrease in the VPF and UPF on small scales, and an increase on large scales. The effect of redshift space distortions is much larger for the simulation with $n = -2$; in this simulation, power exists on all scales up to the box size of the simulation. In redshift space, large scale flows create large underdense regions bounded by caustics. When $n = -2$, the VPF and UPF in redshift space acquire tails which stretch out to areas much larger than the characteristic nonlinearity scale $a_S$.

The VPF is frequently castigated for its sensitivity to the number density $\Sigma$ of galaxies in a survey. The UPF is less sensitive to discreteness effects, but it shares with the VPF the drawback that it treats voids statistically, and doesn't identify individual voids in the large scale structure. In addition, the VPF and UPF in their simplest forms assume that two-dimensional voids are circular. It is frequently useful to be able to detect individual voids in a galaxy survey and to measure their area, their shape, and their alignment relative to the line of sight from the observer. In pursuit of this useful goal, we devised the void-finding algorithm discussed in the previous section, which approximates individual voids as non-overlapping ellipses, within which the number density of galaxies falls below a threshold density, most usefully set equal to $0.2\Sigma(z)$. When voids are defined in this way, it is possible to find how the characteristic size and shape of voids differ in redshift space and real space.

We have shown that the dispersion in void size increases with increasing power on large scales, and is consistent with the idea that $\delta\rho/\rho$ on the scale of the void size expresses itself in void size dispersion. With more work, the void dispersion may be a useful indicator of integrated power.

The size of the largest void in each sample is bigger in redshift space than in real space for all power-law indices $n$ which we studied. However, as Kauffmann & Melott (1992) point out, the maximum void size is a poor choice for defining the characteristic void size in a survey; a better choice is the median or the area-weighted mean void size. The median void size $a_\mathrm{med}$ decreases in going from real space to redshift space when $n = 2$, but increases when $n = 0$ or $n = -2$. The mean void size $a_\mathrm{mean}$ increases in going from real space to redshift space for all three values of $n$. Thus, the statement that voids are larger in redshift space than in real space must be qualified with a statement of how the characteristic void size is determined.

In passing, we note that the literature on voids suffers from the lack of a clear definition, as does that on superclusters, to a lesser degree. It might be appropriate soon for the IAU to consider a formal definition.

With our new improved void-finding algorithm, we are able to ask whether voids in redshift space tend to be stretched out along the line of sight (as they would be if they were expanding outward uniformly in comoving coordinates) or squashed along the line of sight (as they would be if they were contracting uniformly in comoving coordinates). The actual velocity field in a bubbly or spongy distribution will be complex. Examining those voids in each survey with areas $a > a_\mathrm{med}$, we found that these voids do not show a strong tendency to be aligned either with the line of sight or perpendicular to the line of sight. The result of Slezak et al. (1993) – that large voids in the CfA redshift slice are randomly oriented with respect to the line of sight – is thus not saying anything profound about the initial power spectrum of the universe. Random alignments would be found for a wide range of initial spectra. If peculiar velocities fail to cause preferential distortions along the line of sight in redshift space, then any such distortions which are seen in redshift surveys will be due to the cosmological effect described by Alcock & Paczyński (1979). Our future work (Ryden & Melott 1995) will investigate the question of how deep redshift surveys must go before the cosmological distortion in redshift maps becomes reliably detectable.

B.S.R. was supported by NSF grant AST-9357396 and NASA grant NAG 5-2864. A.L.M. was supported by NSF grant AST–9021414 and NASA grant NAGW–3832. Simulations were performed at the National Center for Supercomputing Applications.

|  |  | Void Size | $a_{mean}$ | $a_{sd}$ | $a_{sd}/a_{mean}$ |
|---|---|---|---|---|---|
|  | n=2 | (real) | 1.21e-4 | 0.71e-4 | 0.59 |
|  |  | (redshift) | 1.28e-4 | 0.92e-4 | 0.72 |
| TABLE 1 | n=0 | (real) | 1.06e-4 | 0.87e-4 | 0.82 |
|  |  | (redshift) | 1.67e-4 | 1.57e-4 | 0.94 |
|  | n=-2 | (real) | 3.12e-4 | 7.14e-4 | 2.29 |
|  |  | (redshift) | 16.4e-4 | 22.4e-4 | 1.37 |

Fig. 1.— (a) The distribution of particles in real space for the two-dimensional simulation with power-law index $n = 2$. (b) The distribution of particles in redshift space for the same simulation. Only particles with $z < 0.19$ are included.

Fig. 2.— (a) The distribution of particles in real space for the two-dimensional simulation with power-law index $n = 0$. (b) The distribution of particles in redshift space for the same simulation. Only particles with $z < 0.19$ are included.

Fig. 3.— (a) The distribution of particles in real space for the two-dimensional simulation with power-law index $n = -2$. (b) The distribution of particles in redshift space for the same simulation. Only particles with $z < 0.19$ are included.

Fig. 4.— The void probability function (left hand column) and underdense probability function (right hand column) for each of the simulated universes. Probability is plotted against void area. In each panel, the solid line represents the statistic measured in real space, and the dotted line represents the statistic measured in redshift space. The dashed line is the VPF or UPF for a Poisson distribution of points with the same surface density $\Sigma$ as in the simulations. The UPF has discontinuities because the number threshold is rounded to the nearest integer; this affects the results strongly for $a \lesssim 10^{-5}$

Fig. 5.— (a) The 422 largest underdense voids in the $n = 2$ simulation, as measured in real space; the voids shown occupy 50% of the simulated universe's area. The underdensity threshold for a void is $f = 0.2$. (b) The 446 largest underdense voids in the $n = 2$ simulation, as measured in redshift space; the voids shown occupy 50% of the universe's area.

Fig. 6.— (a) The 585 largest underdense voids in the $n = 0$ simulation, as measured in real space; the voids shown occupy 50% of the simulated universe's area. The underdensity threshold for a void is $f = 0.2$. (b) The 436 largest underdense voids in the $n = 0$ simulation, as measured in redshift space; the voids shown occupy 50% of the universe's area.

Fig. 7.— (a) The 1042 largest underdense voids in the $n = -2$ simulation, as measured in real space; the voids shown occupy 50% of the simulated universe's area. The underdensity threshold for a void is $f = 0.2$. (b) The 184 largest underdense voids in the $n = -2$ simulation, as measured in redshift space; the voids shown occupy 50% of the universe's area.

Fig. 8.— The fraction of the total area of the simulation contained in voids of area $a$ or larger. In the left column, a void is defined as an empty ellipse; in the right column, a void is defined as an ellipse within which the number density of galaxies is equal to $0.2\Sigma(z)$. In each panel, the solid line is the void fraction measured in real space, and the dotted line is the void fraction measured in redshift space. The square symbol represents the area–weighted mean void area on each curve; the triangle the median area void.